\documentclass[a4paper, 10pt]{amsart}

\usepackage[T1]{fontenc}
\usepackage{lmodern}
\usepackage[english]{babel}
\usepackage{amsmath,amssymb,amsthm,amsfonts}
\usepackage{mathrsfs}
\usepackage[all]{xy}
\usepackage{subcaption}
\usepackage{hyperref}
\usepackage{booktabs}
\usepackage{multirow}
\usepackage{float}
\usepackage[backgroundcolor=blue!10]{todonotes}
\hypersetup{
    colorlinks,
    linkcolor={red!50!black},
    citecolor={blue!50!black},
    urlcolor={blue!80!black}
}
\usepackage{algorithm}
\usepackage{algpseudocode} 
\usepackage{enumitem}
\usepackage{graphicx}
\usepackage{subcaption}

\algrenewcommand\algorithmicwhile{\textbf{While}}
\algrenewcommand\algorithmicfor{\textbf{For}}
\algrenewcommand\algorithmicdo{\textbf{Do}}
\algrenewcommand\algorithmicif{\textbf{If}}
\algrenewcommand\algorithmicthen{\textbf{Then}}
\algrenewcommand\algorithmicelse{\textbf{Else}}
\algrenewcommand\algorithmicend{\textbf{End}}
\algrenewcommand\algorithmicreturn{\textbf{Return}}

\theoremstyle{plain}
\newtheorem{lemma}{Lemma}[section]
\newtheorem{proposition}[lemma]{\textbf{Proposition}}

\theoremstyle{definition}
\newtheorem{definition}[lemma]{\textbf{Definition}}

\newtheorem{remark}[lemma]{Remark}

\newcommand{\N}{\mathbb{N}}

\newcommand{\Q}{\mathbb{Q}}

\newcommand{\C}{\mathbb{C}}

\newcommand{\p}{\mathbb{P}}

\newcommand{\mcal}{\mathcal}

\renewcommand{\tilde}{\widetilde}

\begin{document}

\title{Reconstruction of rational ruled surfaces from their silhouettes}
\thanks{This article was published in Journal of Symbolic Computation, Vol 104, May-June 2021, Authors: Matteo Gallet, Niels Lubbes, Josef Schicho, and Jan Vr\v{s}ek, Title: ``Reconstruction of rational ruled surfaces from their silhouettes'', Pages 366--380, \href{https://doi.org/10.1016/j.jsc.2020.08.002}{DOI:10.1016/j.jsc.2020.08.002}, Copyright Elsevier (2020).\\
\copyright\ 2020.  This manuscript version is made available under the CC-BY-NC-ND 4.0 license \url{http://creativecommons.org/licenses/by-nc-nd/4.0/}.}

\author[Matteo Gallet]{Matteo Gallet${}^{\ast, \circ, \diamond, \bullet}$}
\address[MG]{Scuola Internazionale Superiore di Studi Avanzati (SISSA),
Via Bonomea 265, 34136 Trieste, Italy}
\email{mgallet@sissa.it}
\author[Niels Lubbes]{Niels Lubbes${}^{\dagger}$}
\address[NL]{Johann Radon Institute for Computational and Applied 
Mathematics (RICAM), Austrian Academy of Sciences}
\email{niels.lubbes@ricam.oeaw.ac.at}
\author[Josef Schicho]{Josef Schicho${}^{\circ, \diamond}$}
\address[JS]{Research Institute for Symbolic Computation (RISC), Johannes 
Kepler University Linz}
\email{jschicho@risc.jku.at}
\author[Jan Vr\v{s}ek]{Jan Vr\v{s}ek${}^{\ddagger}$}
\address[JV]{University of West Bohemia, Faculty of Applied Sciences}
\email{vrsekjan@kma.zcu.cz}

\thanks{${}^{\ast}$ Corresponding author.}
\thanks{${}^{\circ}$ Supported by the Austrian Science Fund (FWF): W1214-N15, 
Project DK9.} 
\thanks{${}^{\diamond}$ Supported by the Austrian Science Fund (FWF): P26607 and
P31061.}
\thanks{${}^{\bullet}$ Supported by the Austrian Science Fund (FWF): Erwin 
Sch\"odinger Fellowship J4253.}
\thanks{${}^{\dagger}$ Supported by the Austrian Science Fund (FWF): P33003.}
\thanks{${}^{\ddagger}$ Supported by project LO1506 of the Czech Ministry of 
Education, Youth and Sports.}

\begin{abstract}
  We provide algorithms to reconstruct rational ruled surfaces in 
three-dimensional projective space from the ``apparent contour'' of a single 
projection to the projective plane. We deal with the case of tangent 
developables and of general projections to~$\p^3$ of rational normal scrolls. 
In the first case, we use the fact that every such surface is the projection of 
the tangent developable of a rational normal curve, while in the second we 
start by reconstructing the rational normal scroll. In both instances we then 
reconstruct the correct projection to~$\p^3$ of these surfaces by exploiting 
the information contained in the singularities of the apparent contour. 
\end{abstract}

\maketitle

\section{Introduction}

Rational ruled surfaces are rational surfaces that contain a straight line 
through each point. They have been extensively investigated from the point of 
view of computer algebra, see for example~\cite{Chen2001}, \cite{Buse2009}, 
or~\cite{Shen2014}. When we project to~$\p^2$ a rational ruled surface~$S 
\subset \p^3$, we get a finite cover of~$\p^2$ branched along a curve, which is 
the zero set of the discriminant of the equation of~$S$ along the direction of 
projection. In this paper, we study the problem of reconstructing the equation 
of the surface from the discriminant, up to unavoidable projective automorphisms 
that preserve the discriminant. We have already investigated this 
question in~\cite{Gallet2018} for a wider class of surfaces, namely the ones 
admitting at worst ordinary singularities. However, for rational ruled surfaces 
we are able to create a faster algorithm for reconstruction, based on a 
completely different technique from the one in~\cite{Gallet2018}.

We follow the notation from~\cite{Gallet2018} and we define the \emph{contour} 
to be the locus of points on the surface~$S$ whose tangent space passes though 
the center of projection. It consists of the singular locus of the surface and 
of another curve, which we call the \emph{proper contour}. The projection of the 
contour is the \emph{silhouette} of~$S$, which splits into the \emph{singular 
image} and the \emph{proper silhouette}. Our goal is then, starting from the 
knowledge of the singular image and of the proper silhouette, to reconstruct the
surface~$S$, up to projective automorphisms that preserve the center of projection
and the lines through it.

In the remainder of the introduction, we discuss the organization of the paper.

A special subcase of rational ruled surfaces is the one of developables, i.e.,  
surfaces that are either a cone over a plane curve or the union of the 
tangents of a space curve. In Section~\ref{tangent_developable} we consider the 
latter case. The main idea is that the projection of the 
space curve appears as a component with multiplicity three in the discriminant. 
Our task is to lift this projected curve back to space; we do so by 
exploiting the information contained in particular singularities of the 
projection of the nodal curve of the surface. In this subcase, the proper silhouette
consists of a union of lines, which are projections of special lines 
of the surface. 

In Section~\ref{general_projection} we deal with general projections to~$\p^3$ 
of rational normal scrolls. In this case, we first identify the rational 
normal scroll and construct a projection from it to~$\p^2$ having the proper 
silhouette as branching locus. Secondly, we project the rational normal 
scroll to~$\p^3$ so that we obtain a double curve as prescribed by the singular 
image. Notice that the two-dimensional picture of a ruled surface tells more than the 
picture of an arbitrary surface: the lines of the surface map to tangents 
to the apparent contour, hence we already get (for free!)  projections 
of all lines of the surface. This is an advantage which we try to keep in 
our algorithm. The rationality of the given ruled surface implies then that the 
apparent contour is itself rational. There are well-known algorithms 
to produce a parametrization, and such a parametrization can also be 
used to parametrize the set of all tangents, namely its dual 
curve. The $\mu$-basis of the parametrization, which was defined and 
studied in the context of the implicitization problem by \cite{Cox1998},
gives rise to a rational normal scroll surface, embedded in a 
projective space of higher dimension. The ruled surface that we wish to 
compute is then simply a projection of this rational scroll, and the 
most expensive computations go into figuring out this final projection.

We conclude the paper by recalling in Section~\ref{parametrization}, for the 
benefit of the reader, a known algorithm for the parametrization of curves which 
we implement \emph{ad hoc} in our algorithm, since currently available 
general-purpose algorithms for parametrizations of plane curves do not exploit 
the special structure of the curves we deal with. This determines a relevant 
speed-up of our algorithm, since the parametrization of a particular 
planar curve constitutes its computational bottleneck.

An implementation in Maple of the algorithms developed in this paper is 
available at
\begin{center}
 \url{https://www.risc.jku.at/people/jschicho/pub/ChisiniRuled.mpl}.
\end{center}

\section{Tangent developable surfaces}
\label{tangent_developable}

Among ruled surfaces, developable surfaces can be characterized as follows. A 
line of a ruled surface is called \emph{torsal} if all tangent 
planes at all smooth points of the line are equal; a \emph{developable surface} 
is then a ruled surface such that all lines of the surface are torsal.

\begin{figure}
\centering
 \includegraphics[width=.5\textwidth]{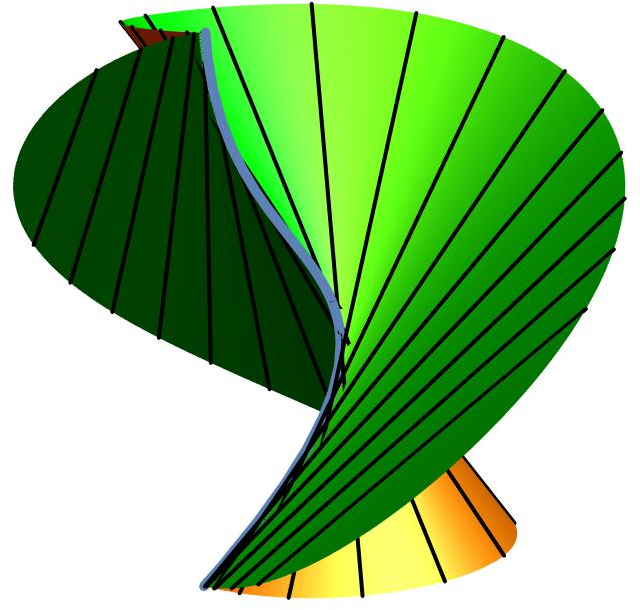}
 \caption{The tangent developable surface of a twisted cubic, highlighted as a 
blue thick curve.}
 \label{figure:tangent_developable}
\end{figure}

A developable surface is either a \emph{tangent 
developable}, i.e., the union of all tangents of a 
space curve (Figure~\ref{figure:tangent_developable}), or a cone over a planar 
curve --- see \cite[Chapter~IV, Section~30]{Hilbert1952} and 
\cite[Theorem~0]{Ushakov1999}, or \cite[Proposition~2.12]{Arrondo} for a proof 
in terms of algebraic geometry. In this section, we recognize general rational 
tangent developables from 
their silhouettes with respect to general projections. The case of cones is 
equivalent to the recognition of a planar algebraic curve from its branching 
points with respect to a projection to the projective line.

Let $d \ge 3$. Every tangent developable of a rational curve of degree~$d$ is a 
projection of the tangent developable $T_d \subset \p^d$
of the rational normal curve of degree~$d$ in~$\p^d$. The surface $T_d$ has 
degree $2d - 2$ and admits a parametrization $\C^2 \longrightarrow T_d$ of the 
form
\begin{align*}
 (s,t) &\mapsto \bigl(1 : (t+s) : (t^2 + 2st) : (t^3 + 3t^2 s): \cdots: (t^d + 
dt^{d-1} s) \bigr) \,. 
\end{align*}
The surface~$T_d$ contains the rational normal curve as a cuspidal 
singularity, namely locally analytically the surface around such a curve 
looks like a cylinder over a plane cusp. Hence tangent developables have a 
whole curve composed of cuspidal singularities, which is the projection of the 
rational normal curve in~$T_d$. We call this curve the \emph{cuspidal curve}, 
and we notice that it is a rational curve. Tangent developables may have other 
singularities, coming from the fact that they are projections of~$T_d$, such as 
self-intersecting curves. Locally analytically around a point of a 
self-intersection curve, the tangent developable has two branches that 
intersect transversally. This is why we call these curves of singularities 
\emph{nodal curves}. 

\begin{figure}
 \centering
 \begin{tabular}{ccc}
 \begin{subfigure}[t]{0.35\textwidth}
  \includegraphics[height=\textwidth]{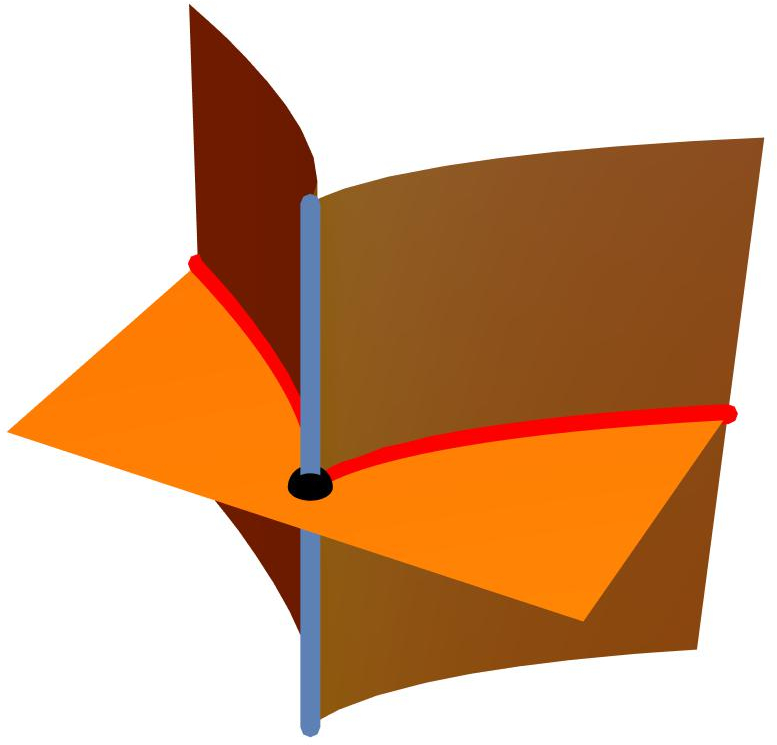}
\caption{Cuspidal intersection between the nodal curve (red) and the cuspidal 
curve (blue).}
\label{figure:cuspidal_intersection}
 \end{subfigure}
  & \quad &
 \begin{subfigure}[t]{0.35\textwidth}
  \includegraphics[height=\textwidth]{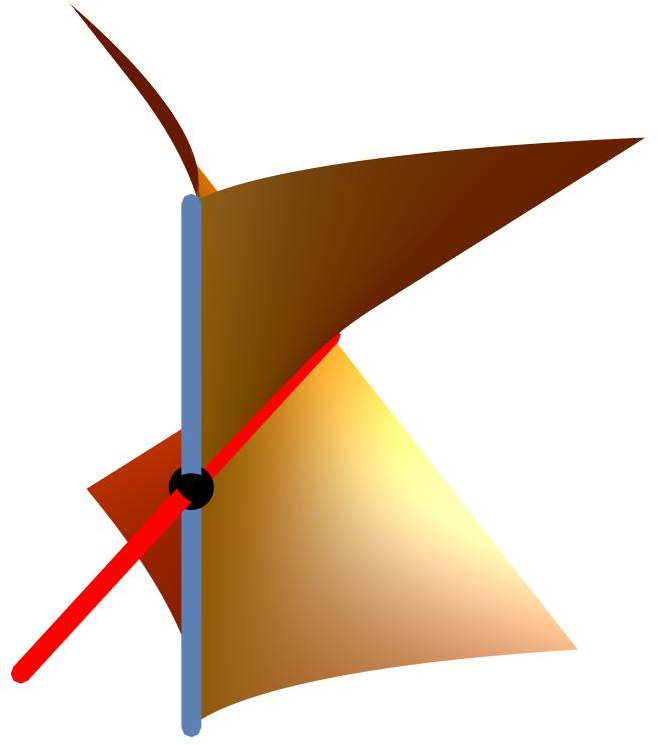}
\caption{Cuspidal pinch point, a transversal intersection between the nodal 
curve (red) and the cuspidal curve (blue).}
  \label{figure:transverse_intersection}
 \end{subfigure}
 \end{tabular}
 \caption{Two possible types of intersection between the cuspidal and the nodal 
curve of a tangent developable.}
\end{figure}

Here we describe the tangent developables in~$\p^3$ we intend to recognize: we 
prescribe their singularities having in mind the situation of a general 
projection of~$T_d$ to~$\p^3$. 
We call a tangent developable in~$\p^3$ \emph{good} if it satisfies the 
following conditions:
\begin{itemize}
 \item the only singularities of the tangent developable are contained in the 
cuspidal curve, and in the nodal curve;
 \item the cuspidal curve is smooth and irreducible;
 \item the nodal curve is irreducible and has only ordinary triple points, or 
 singular points as described in the next item;
 \item the points of intersection of the nodal curve and the cuspidal curve 
belong to one of the following two types:
\begin{itemize}
 \item points for which the local analytic equation of the surface at the 
point is equivalent to $(x^2-y^3)z=0$; in this local equation, the point is the 
origin, the cuspidal curve is $x=y=0$ and the nodal curve is $x^2-y^3=z=0$; as 
the local equation shows, the nodal curve has a cusp at the intersection point 
(see 
Figure~\ref{figure:cuspidal_intersection});
 \item points that are a transversal intersection of the nodal and the cuspidal 
curves at a ``cuspidal pinch point''; the local analytic equation at such a 
point is equivalent to $z^2 y^3 - x^2 = 0$ (see 
Figure~\ref{figure:transverse_intersection});
\end{itemize}
 \item there are exactly $4(d-3)$ cuspidal pinch points.
\end{itemize}

Now we examine the silhouette with respect to a general projection to~$\p^2$ of 
a good tangent developable surface. In the algorithm we are going to develop 
(Algorithm \texttt{ReconstructTangentDevelopable}), we use the knowledge of the 
two components of the silhouette to reconstruct the tangent developable.

\begin{definition}
 Let $S \subset \p^3$ be a good tangent developable surface. The image of 
the cuspidal curve of~$S$ under a 
projection to~$\p^2$ is called  the \emph{cuspidal image}; the image of the 
nodal curve of~$S$ is called the \emph{nodal image}.
\end{definition}

\begin{proposition}
If $S \subset \p^3$ is a good tangent developable surface, then the 
discriminant of a general projection\footnote{The \emph{discriminant} of a 
projection $\pi \colon S \longrightarrow \p^2$ is the polynomial defining the 
locus of points $P \in \p^2$ such that the preimage of~$P$ under~$\pi$ is 
tangent to~$S$. If, in coordinates, the projection is given by the map 
$\pi \colon (x:y:z) \mapsto (x:y)$ and $F = 0$ is the equation of~$S$, then the 
discriminant of~$\pi$ is the discriminant of~$F$ with respect to~$z$.} of~$S$ 
to~$\p^2$ has the following factors:
\begin{itemize}
\item a factor of multiplicity three, whose zero set is the 
cuspidal image;
\item a factor of multiplicity two, whose zero set is the nodal image;
\item several linear factors of multiplicity one, which are images of tangent 
planes passing through the center of projection; their zero set is the proper 
silhouette of the projection.
\end{itemize}
\end{proposition}

\begin{proof}
A general projection of a cusp is a triple zero of the discriminant, hence the 
cuspidal image is a triple component. Similarly, a general projection of a 
node is a double zero of the discriminant, hence the nodal image is a double 
component. Both these two results follow from a local analysis of the 
situation. In fact, locally around a general point of the cuspidal curve,
we can take coordinates so that the equation of the surface is $z^2-x^3$ and 
the projection is along~$z$, so the discriminant is~$-4x^3$; similarly, locally
around a general point of the nodal curve, we can take the equation of the 
surface to be $(z-x)(z+x)$, and the projection along~$z$, so that the discriminant is~$-4x^2$.
Since all lines are torsal, namely all the tangent planes at points 
of one of these lines coincide, the proper contour is composed of lines, 
and their images determine the linear factors of the discriminant.
\end{proof}

The lines in the proper silhouette are inflection tangent lines of the cuspidal 
image. The intersections of the cuspidal image and the nodal image are either 
cusps of the nodal image or transverse intersections coming from transverse 
intersections of nodal and cuspidal curve, or from pairs of distinct point, each 
on one of the two curves, collapsed by the projection. Apart from these 
intersections, and from triple points on the nodal image, we allow the cuspidal 
image and the nodal image to have only ordinary double points. 

The recognition problem of a tangent developable reduces to the 
construction of the cuspidal curve from the factors of the discriminant. The 
cuspidal curve is the projection of a rational normal curve of the 
same degree~$d$. This projection is given by four 
polynomials of degree~$d$ of which we already know three, namely those that 
define the projection to the cuspidal image in~$\p^2$. In order to find 
the fourth polynomial, we have to use the nodal image. Notice, in fact, that 
the inflection lines do not add any information: any choice of a fourth 
polynomial would lead to the same linear factors of the discriminant.

The basic idea of Algorithm~\texttt{ReconstructTangentDevelopable} is to collect 
linear conditions for the fourth polynomial derived from cuspidal pinch 
points. Their projections to~$\p^2$ are transversal 
intersections of the nodal image with the cuspidal image. Hence, we try every 
possible subset of cardinality~$4(d-3)$ of the set of all transversal 
intersections. Notice that if the cuspidal curve is defined 
over~$\Q$, then the images of the cuspidal pinch points are all conjugated 
over~$\Q$, so it is easy to extract them from the set of all intersections. 
Suppose that the three polynomials giving the parametrization of the cuspidal 
image are $H_0$, $H_1$ and~$H_2$. Then the fourth polynomial~$H_3$ can be found 
as follows: we make a symbolic ansatz for its coefficients and compute the 
determinant of the matrix
\begin{equation}
\label{eq:cuspidal}
 \begin{pmatrix}
  \frac{\partial^3 H_0}{\partial t_0^3} & 
  \frac{\partial^3 H_0}{\partial t_0^2t_1} &
  \ldots &
  \frac{\partial^3 H_0}{\partial t_1^3} \\
  \vdots & \vdots & & \vdots \\
  \frac{\partial^3 H_3}{\partial t_0^3} & 
  \frac{\partial^3 H_3}{\partial t_0^2t_1} &
  \ldots &
  \frac{\partial^3 H_3}{\partial t_1^3} \\
 \end{pmatrix} \, .
\end{equation}
This determinant vanishes at the parameters corresponding to 
images of the cuspidal pinch points (see Remark~\ref{remark:multiplicities}). 
Imposing this vanishing for all points of a subset of cardinality $4(d-3)$ of 
transversal intersections of the nodal image with the cuspidal image provides 
linear conditions for the coefficients of~$H_3$. We show 
in Lemma~\ref{lemma:tangent_developable} that the space of solutions of these 
linear equations is $4$-dimensional. This means that the choice of~$H_3$ is 
unique up to linear combinations of~$H_0$, $H_1$, and~$H_2$. In turn, this 
clarifies that the reconstruction of the cuspidal curve (and hence of the 
tangent developable) performed by 
Algorithm~\texttt{ReconstructTangentDevelopable} is unique up to projective 
transformations of~$\p^3$ that preserve the silhouette.

\begin{algorithm}[ht]
\caption{\texttt{ReconstructTangentDevelopable}}
\begin{algorithmic}[1]
  \Require The equation of a rational curve $C \subset \p^2$, the image of the 
cuspidal curve of a good tangent developable, and of another curve~$D$, the 
image of the nodal curve.
  \Ensure The parametrization of the cuspidal curve of the tangent developable 
$S \subset \p^3$.
  \Statex
  \State {\bfseries Compute} a parametrization $(H_0: H_1: H_2)$ of~$C$.
  \State {\bfseries Formulate} a symbolic ansatz for the coefficients of~$H_3$ 
and compute the determinant of the matrix in Equation~\eqref{eq:cuspidal}.
  \State {\bfseries Select} a set $T$ of $4(d-3)$ transverse intersections 
of~$C$ and~$D$ (the candidates for the images of cuspidal pinch points). 
  \For{each point $x$ of~$T$}
  \State {\bfseries Evaluate} the determinant at~$x$ and collect the linear 
equations in the coefficients of~$H_3$. 
  \EndFor
  \State {\bfseries Solve} the system of linear equations and obtain $H_3$.
  \State \Return the parametrization $(H_0: H_1: H_2: H_3)$.
\end{algorithmic}
\end{algorithm}

We would like to point out that Algorithm 
\texttt{ReconstructTangentDevelopable} relies on nontrivial ``basic'' 
operations in computational algebraic geometry, as computing a parametrization 
of a rational curve, or computing and manipulating the intersection of two 
plane curves. These are, in general, difficult independent problems for which 
there is a vast literature, and that are objects of active research; we 
consider them as ``building blocks'' of our algorithm.

Notice that, in order to perform Step~$3$, one may plug the parametrization 
of~$C$ into the equation of~$D$, hence obtaining an equation in a single 
variable. Then this equation may be solved (symbolically, which may need the 
introduction of algebraic numbers, or numerically), and the solutions provide 
the points of intersection of~$C$ and~$D$.

\begin{remark}
\label{remark:multiplicities}
 Consider a nondegenerate rational 
curve~$C$ of degree~$d$ in~$\p^n$, and a point $P \in C$, which we can suppose 
to be~$(1: 0: \ldots: 0)$. A local parametrization of~$C$ 
around~$P$ is of the form $\bigl( f_0(t): f_1(t): \ldots: f_n(t) \bigr)$ with 
$\mathrm{ord}(f_i) = \alpha_i$. We may assume that $0 = \alpha_0 < \alpha_1 < 
\ldots < \alpha_n$ up to linear changes of coordinates. We say that $P$ is 
\emph{special} if $\alpha_n > n$. The \emph{multiplicity} of a special point is 
defined as $\sum_{i=0}^n \alpha_i - \binom{n+1}{2}$, and one has that the sum 
of 
multiplicities of special points of~$C$ equals $(n+1)(d-n)$ (see 
\cite[Definition~3.4 and Lemma~3.6]{GalletSchicho2017}). A general rational 
curve~$C$ of degree~$d$ in~$\p^3$ has exactly~$4(d-3)$ special points of 
multiplicity~$1$ (see \cite[Lemma~3.5]{GalletSchicho2017}), and a local 
parametrization of~$C$ at these points is of the form $(1 + \ldots: t + 
\ldots: t^2 + \ldots: t^4 +\ldots)$. These points determine the cuspidal pinch 
points of tangent developable of~$C$.
\end{remark}

\begin{lemma}
\label{lemma:tangent_developable}
 The dimension of the solution space to the linear system in Algorithm 
\texttt{ReconstructTangentDevelopable} is~$4$. 
\end{lemma}
\begin{proof}
 The dimension is at least $4$ since we suppose that we start from a projection 
of a tangent developable. In particular, there exists a rational curve~$C$ of 
degree $d$ in $\p^3$ parametrized by $(H_0: H_1: H_2: H_3)$ where the $H_i$ are 
linearly independent elements in the solution space. By assumption, the 
curve~$C$ admits exactly $4(d-3)$ special points of multiplicity~$1$: in fact, 
if $C$ has a special point~$P$ of multiplicity~$1$, we see from its local 
parametrization that the tangent developable of~$C$ has a cuspidal pinch point 
at~$P$. Suppose, by 
contradiction, that the dimension of the solution space is at least~$5$. This 
implies that there exists a rational curve~$C'$ of degree~$d$ in~$\p^4$ 
projecting to~$C$. The curve~$C'$ has at least $4(d-3)$ special points that 
remain special after projection to $\p^3$. Therefore, these points must be of 
multiplicity at least~$2$. This follows from the fact that special points of 
multiplicity~$1$ project to non-special points, since any 
multiplicity~$1$ point on the curve~$C'$ has local parametrization of type 
$(1: t + \ldots: t^2 + \ldots: t^3 + \ldots: t^5 + \ldots)$; these points 
project to points in~$\p^3$ around which the parametrization is $(1: t + \ldots: 
t^2 + \ldots: t^3 + \ldots)$, and so they are non-special. The sum of 
multiplicities of special points in~$C'$ must give $5(d-4)$, but this is a 
contradiction, since $2 \cdot 4(d-3) > 5(d-4)$ for $d \geq 3$. 
\end{proof}

\begin{remark}
 Notice that there is no nodal curve in the tangent developable of a twisted 
cubic, and thus no special points in its cuspidal curve. However, 
Lemma~\ref{lemma:tangent_developable} is trivially true for $d = 3$ since the 
space of univariate polynomials of degree at most~$3$ is $4$-dimensional.
\end{remark}

\section{General projections of rational normal scrolls}
\label{general_projection}

In this section, we provide a reconstruction algorithm for rational ruled 
surfaces that admit particularly simple singularities. Recall from the 
introduction that, given a projection~$S \longrightarrow \p^2$, we call 
\emph{contour} the locus of points on $S$ whose tangent plane passes 
though the center of projection. The contour is the union of singular locus 
of the surface and of the \emph{proper contour}. The image of the 
contour under the projection is the \emph{silhouette}; it is constituted 
of the \emph{singular image} (the image of the singular locus) and the 
\emph{proper silhouette} (the image of the proper contour). In this section, we 
consider \emph{good} rational ruled surfaces $S \subset \p^3$ and \emph{good} 
projections $S \longrightarrow \p^2$, namely we ask that:
\begin{itemize}
 \item $S$ has at most ordinary singularities: an 
irreducible self-intersection curve, self-intersection triple points, and pinch 
points;
 \item the proper silhouette has only nodes and cusps;
 \item the singular image has only nodes and ordinary triple points;
 \item the proper contour is irreducible and projects birationally to the 
proper silhouette;
 \item the singular curve projects birationally to the singular image.
\end{itemize}
Recall that every rational ruled surface is a projection of a rational normal 
scroll. Our assumptions imply that the restriction of this projection to its 
ramification locus is generically injective.  
Notice that all our assumptions are fulfilled when we consider projections from 
general centers (both when we 
project from the rational normal scroll to~$\p^3$, and when we project from 
the surface~$S$ to~$\p^2$). In particular, irreducibility of the singular curve 
is Franchetta's Theorem (see~\cite[Theorem~6]{Mezzetti1997}).
The fact that good rational ruled surfaces have ordinary singularities implies 
that they have finitely many torsal lines, each passing through exactly one 
pinch point. 

We divide the reconstruction process in two steps: first, we determine the 
rational normal scroll of which the surface is a projection, together with the 
projection from this rational normal scroll to~$\p^2$, by computing a 
parametrization of the dual of the proper silhouette, which is a rational 
curve; second, we construct the projection map from the rational normal scroll 
to the surface in~$\p^3$.
\begin{figure}
 \begin{tabular}{ccc}
  \includegraphics[width=.45\textwidth]{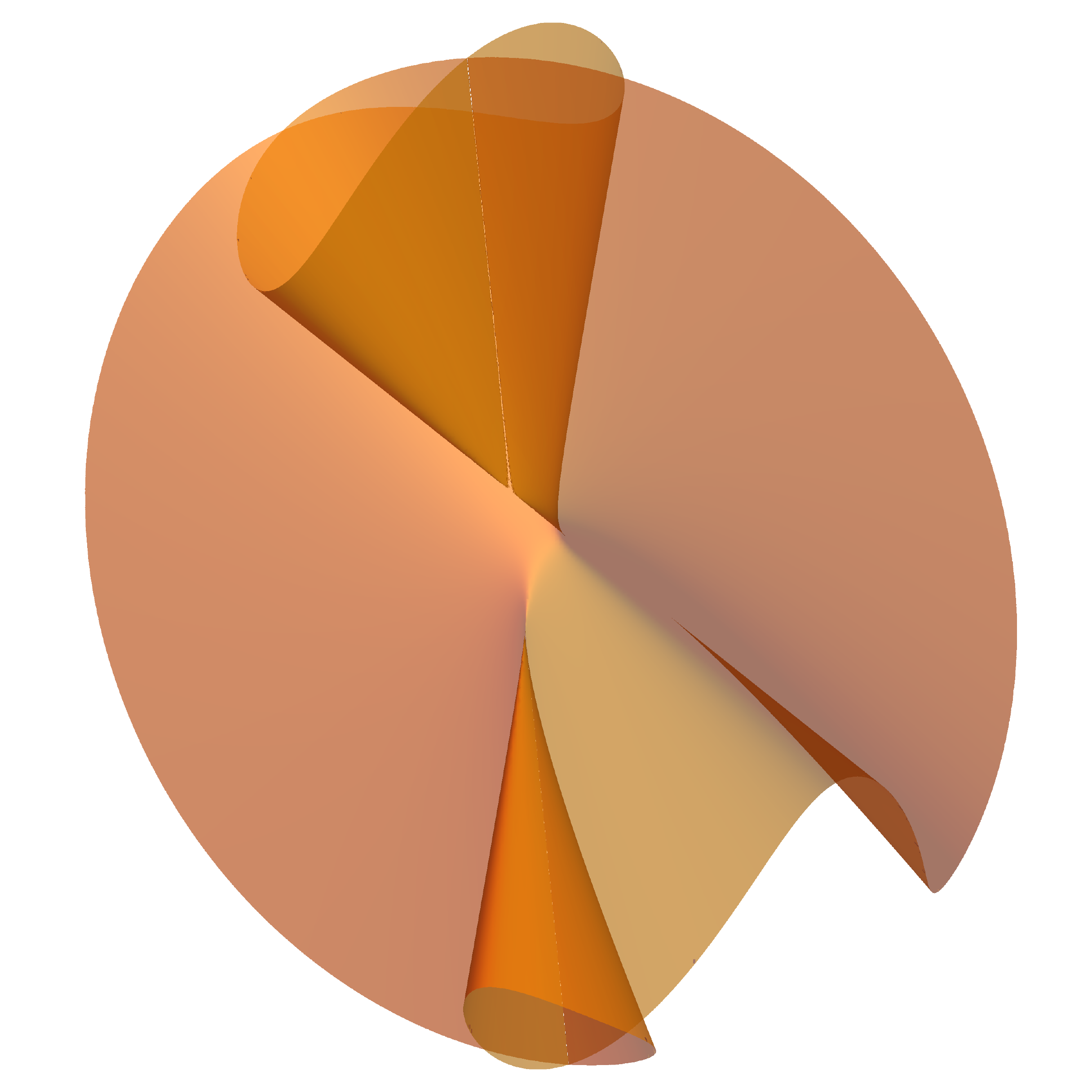}
  & & 
  \includegraphics[width=.45\textwidth]{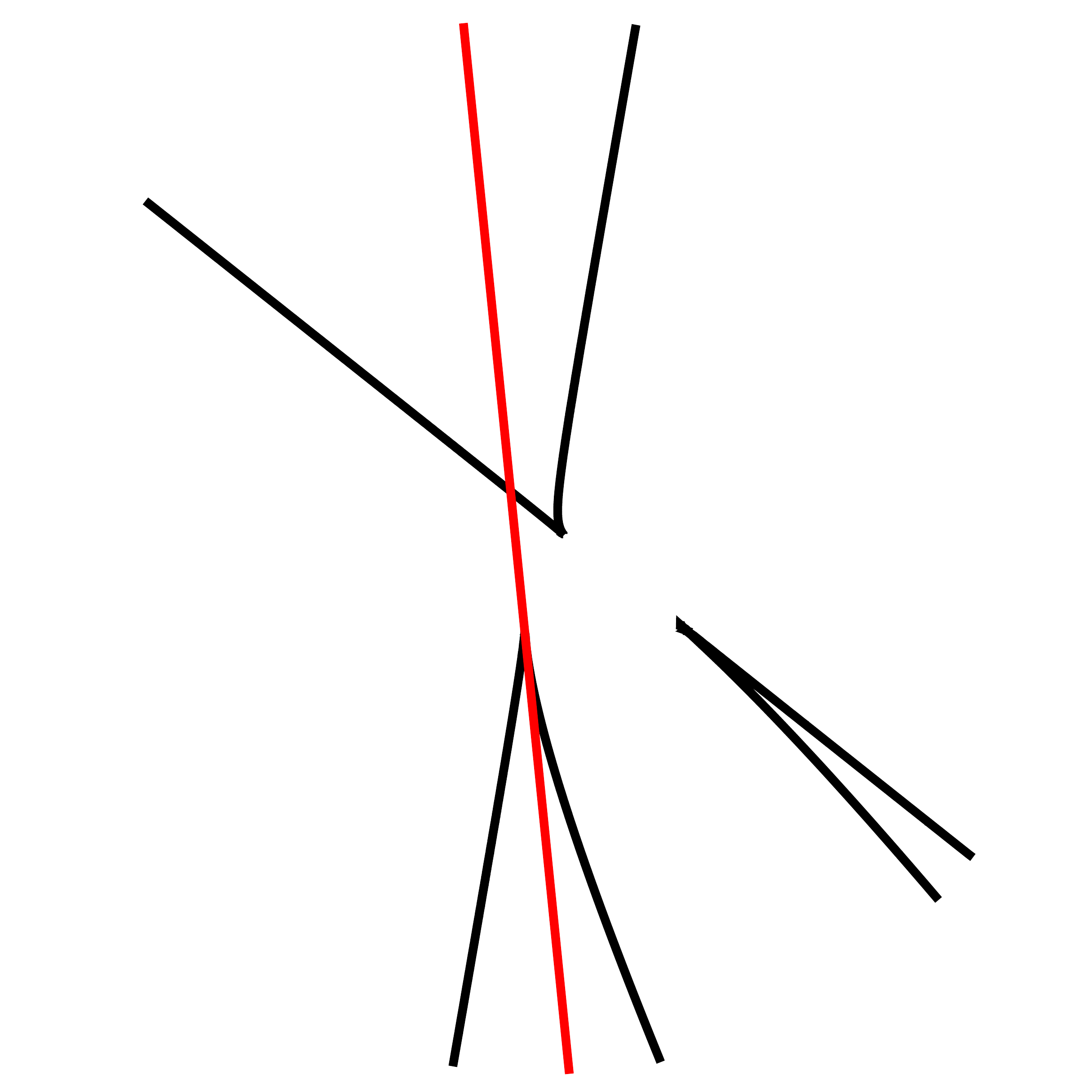}
 \end{tabular}
 \caption{A singular ruled cubic and its silhouette, constituted of the 
singular image (straight line in red) and the proper silhouette (in black).}
\end{figure}
To explain how the first part of the reconstruction works, we recall 
some facts about rational normal scrolls.

For our purposes, we use the description of \emph{rational normal scrolls} 
provided by \cite[Example~2.3.16]{Cox2011}: given two natural numbers $d_1, d_2 
\in \N$, the rational scroll~$\Sigma_{d_1, d_2}$ is the Zariski closure of the 
image of the map
\[
 p_{d_1, d_2} \colon
 \bigl( \C^{\ast} \bigr)^2 \longrightarrow \p^{d_1 + d_2 + 1}, 
 \qquad
 (s,t) \mapsto (1: t: \dotsc: t^{d_2}: s: st: \dotsc: st^{d_1}) \, .
\]
In this way, a linear projection $\rho \colon \Sigma_{d_1, d_2} \longrightarrow 
\p^2$ can be identified with two vectors of polynomials $Q_1, Q_2 \in \C[t]^3$ 
of degrees~$d_1$ and~$d_2$ so that the map $r \colon \bigl( \C^{\ast} \bigr)^2 
\longrightarrow \p^2$
\begin{equation}
\label{eq:toric}
 (s,t) \mapsto 
  \bigl( 
   Q_{20}(t) + s\,Q_{10}(t): Q_{21}(t) + s\,Q_{11}(t): Q_{22}(t) + s\,Q_{12}(t)
  \bigr)
\end{equation}
fits into the commutative diagram
\[
 \xymatrix{
  \Sigma_{d_1, d_2} \ar[r]^{\rho} & \p^2 \\
  \bigl( \C^{\ast} \bigr)^2 \ar[u]^{p_{d_1, d_2}} \ar[ur]_{r}
 }
\]
Since the projection $\rho \colon \Sigma_{d_1, d_2} \longrightarrow \p^2$ is a 
map between smooth varieties, the notions analogous to the ones of contour and 
silhouette for projections $S \longrightarrow \p^2$ are usually called, in this 
case, the \emph{ramification locus} and the \emph{branching locus}, 
respectively. Hereafter, we collect some pieces of information about the 
branching locus of such a projection. The following result is 
well-known, but we report the proof, since we could not find a reference 
asserting exactly the fact we need.

\begin{lemma}
\label{lemma:branching_envelope}
 Let $\rho \colon \Sigma_{d_1, d_2} \longrightarrow \p^2$ be a
projection from a rational normal scroll whose restriction to the ramification 
locus~$R \subset \Sigma_{d_1, d_2}$ is generically injective. Then every line 
in~$\Sigma_{d_1, d_2}$ is projected by~$\rho$ to a tangent line of the branching 
locus.
\end{lemma}

\begin{proof}
Let $L \subset \Sigma_{d_1, d_2}$ be a line. By hypothesis, we know that 
$\rho_{|_{L}}$ and~$\rho_{|_{R}}$ are both generically injective because any 
projection whose center is not in~$\Sigma_{d_1, d_2}$ is generically injective 
on the lines of the surface. Let $p \in L\cap R$ be a smooth point of~$R$, then 
$\rho(L)$ is tangent to~$\rho(R)$ at~$\rho(p)$. In fact, by 
definition, the tangent plane of~$\Sigma_{d_1, d_2}$ at~$p$ intersects the 
center of projection. The tangent line of~$R$ at~$p$ and~$L$ are both contained 
in this tangent plane. The projection either collapses the tangent line of~$R$. 
In the first case, since $\rho_{|_{R}}$ is generically injective, and there are no 
birational maps between smooth curves collapsing tangent vectors, it follows 
that $\rho(p)$ is singular in~$\rho(R)$. Therefore the statement is trivially 
true. In the second case, both $L$ and the tangent line of~$R$ are projected to a single line 
in~$\p^2$, which equals~$\rho(L)$. Hence $\rho(L)$ is tangent to~$\rho(R)$ at~$\rho(p)$.
\end{proof}

The next proposition provides the first part of the reconstruction algorithm. 
We are going to use the concept of $\mu$-bases, introduced by Cox, Sederberg 
and Chen in~\cite{Cox1998}: these are particular sets of generators of the 
module of syzygies of a parametrization of a curve or of a surface; we refer to 
the original paper for a more precise definition and for their properties.

\begin{proposition}
\label{proposition:correctness}
 Let $B \subset \p^2$ be the branching locus of a projection $\Sigma_{d_1, 
d_2} \longrightarrow \p^2$ whose restriction to the ramification locus is 
generically injective. Let $(Q_1, Q_2)$ be a $\mu$-basis of a parametrization 
of~$\check{B}$, the dual curve of~$B$. Then the projection $\rho \colon 
\Sigma_{d_1, d_2} \longrightarrow \p^2$ induced by $(Q_1, Q_2)$ has~$B$ as 
branching locus. 
Moreover, every general projection $\tilde{\rho} \colon \Sigma_{\delta_1, 
\delta_2} \longrightarrow \p^2$ having~$B$ as branching locus is projectively 
equivalent to~$\rho$ over~$B$, namely there exists a projective automorphism 
$\alpha \colon \Sigma_{\delta_1, \delta_2} \longrightarrow \Sigma_{d_1, d_2}$ 
such that $\alpha$ fixes all linear spaces of dimension $d_1+d_2-1$ through 
the center of projection, which has dimension~$d_1+d_2-2$.
\end{proposition}
\begin{proof}
We first show that $B$ is the branching locus of the map $\rho \colon 
\Sigma_{d_1, d_2} \longrightarrow \p^2$ obtained from $(Q_1, Q_2)$. By the 
properties of $\mu$-bases (see \cite[Theorem~2]{Song2009}), the vector $Q_1 
\times Q_2$ gives a parametrization of~$\check{B}$. This implies that the 
projection~$\rho$ sends the lines of~$\Sigma_{d_1, d_2}$ to the family of 
tangent lines of~$B$. In fact, the Pl\"ucker coordinates of the line 
between~$Q_1(t)$ and~$Q_2(t)$ are given by $\bigl(Q_1 \times Q_2\bigr)(t)$. By 
Lemma~\ref{lemma:branching_envelope}, the curve~$B$ is the branching locus 
of~$\rho$. Suppose that $\phi \colon \Sigma_{\delta_1, \delta_2} \longrightarrow 
\p^2$ is another general projection having~$B$ as branching locus. 
Let $\tilde{Q}_1, \tilde{Q}_2, \in \C[t]^3$ be the two vectors of polynomials 
defining~$\phi$ as explained in Equation~\eqref{eq:toric}. By 
Lemma~\ref{lemma:branching_envelope}, the images of the lines of 
$\Sigma_{\delta_1, \delta_2}$ are lines tangent to $B$. Hence, by the same 
argument as before, $\tilde{Q}_1 \times \tilde{Q}_2$ is a parametrization 
of~$\check{B}$. Thus $(\tilde{Q}_1, \tilde{Q}_2)$ is a $\mu$-basis of this 
parametrization. By the uniqueness of $\mu$-bases, it follows that $\delta_1 = 
d_1$ and $\delta_2 = d_2$, and that $\phi$ and $\rho$ differ by an automorphism 
over~$B$.
\end{proof}

Proposition~\ref{proposition:correctness} tells us that, in order to reconstruct
the rational normal scroll of which the desired surface in~$\p^3$ is a projection
it is enough to compute a $\mu$-basis of a parametrization of the dual of
the proper silhouette.

This concludes the first part of the reconstruction process, and proves the 
correctness of Algorithm \texttt{ReconstructRationalScroll}: there, instead of 
computing the map~$\rho$, we compute the map~$r$ as in 
Equation~\eqref{eq:toric}, which provides the same information. In fact, the 
proper silhouette of a projection $S \longrightarrow \p^2$ is the branching 
locus of the corresponding projection $\Sigma_{d_1, d_2} \longrightarrow \p^2$.

\begin{algorithm}
\caption{\texttt{ReconstructRationalScroll}}
\begin{algorithmic}[1]
  \Require A curve $B \subset \p^2$, the proper silhouette of a 
projection~$S$ of a rational normal scroll~$\Sigma_{d_1, d_2}$
whose restriction to the ramification locus is generically injective.
  \Ensure Two numbers $d_1, d_2 \in \N$ and a map $r \colon (\C^{\ast})^2 
\longrightarrow \p^2$ as in Equation~\eqref{eq:toric}, whose branching locus 
is~$B$.
  \Statex
  \State {\bfseries Compute} the dual curve~$\check{B}$ of~$B$.
  \State {\bfseries Parametrize} the curve~$\check{B}$.
  \State {\bfseries Compute} a $\mu$-basis $(Q_1, Q_2)$ of the parametrization 
of~$\check{B}$.
  \State {\bfseries Set} $(d_1, d_2) = \bigl( \deg(Q_1), \deg(Q_2) \bigr)$.
  \State {\bfseries Set} $r$ to be the map 
  \[ 
   (s,t) 
   \mapsto 
   \bigl( 
    Q_{20}(t) + s\,Q_{10}(t): Q_{21}(t) + s\,Q_{11}(t): Q_{22}(t) + 
s\,Q_{12}(t) 
    \bigr) \, .
  \]
  \State \Return $d_1$, $d_2$ and $r$.
\end{algorithmic}
\end{algorithm}

As a consequence of Algorithm \texttt{ReconstructRationalScroll}, we obtain a 
characterization of proper silhouettes of good projections $S \longrightarrow 
\p^2$, where $S$ is a good rational ruled surface.

\begin{proposition}
  Proper silhouettes of good projections $S \longrightarrow \p^2$ of 
good rational ruled surfaces~$S$ are rational plane curves with maximal number 
of cusps for a given degree and only ordinary nodes as the other singularities. 
\end{proposition}
\begin{proof}
Let $B$ be a proper silhouette of a good projection $S \longrightarrow \p^2$, 
where $S$ is a good rational ruled surface. We know that the degree of the 
proper silhouette~$B$ is $n = 2d - 2$, where $d = \deg(S)$. We can obtain this 
number by considering the degree of the silhouette, which is $d(d-1)$, and 
subtracting from it twice the degree of the singular image, which is 
$\frac{1}{2} (d-1)(d-2)$ (see \cite[Section~6]{Piene2005}).
If we denote by~$\delta$ the number of nodes of~$B$ and by~$\kappa$ the number 
of cusps of~$B$, then from the rationality of~$B$ and the Pl\"ucker formulas we 
get
\[
 \frac{(n-1)(n-2)}{2} - \delta - \kappa = 0 
 \quad \text{and} \quad 
 n^{\ast} = n(n-1) - 2\delta - 3\kappa, 
\]
where $n^{\ast}$ is the degree of the dual of~$B$. Since the dual of~$B$ is in 
fact a plane section of the dual of~$S$, and the dual of a ruled surface is a 
ruled surface of the same degree, we conclude that $n^{\ast} = d$. These two 
equations imply that $\kappa = \frac{3}{2}(n-2)$. On the other hand, 
we know that for a rational curve of degree~$n$ the number of cusps must be less 
than or equal to~$\frac{3}{2}(n-2)$ (see \cite[Section~5]{Lefschetz1913} and 
\cite[Section~4]{Hollcroft1937}). Hence proper silhouettes of rational ruled 
surfaces are rational curves with maximal number of cusps for a given degree. 

Conversely, if we are given a rational curve of degree~$n$, where $n$ is of the 
form $2d - 2$ for some~$d$, having $\frac{3}{2}(n-2)$ ordinary cusps and only 
ordinary nodes as the other singularities, then we can apply Algorithm 
\texttt{ReconstructRationalScroll} to it, thus showing that such a curve is the 
silhouette of the projection of a rational ruled surface whose restriction to 
the ramification locus is generically injective. 
\end{proof}

Now that, recalling Equation~\eqref{eq:toric}, we reconstructed the map $\rho 
\colon \Sigma_{d_1, d_2} \longrightarrow \p^2$ via Algorithm 
\texttt{ReconstructRationalScroll}, we proceed by recovering the projection 
$\Sigma_{d_1, d_2} \longrightarrow S$. 
To do so, we need to use the information provided by the singular image, namely 
the projection on~$\p^2$ of the singular locus of~$S$. We formulate two 
different algorithms to construct this projection: one (Algorithm 
\texttt{CollapseMates}) imposes the projection to collapse pairs of points in 
$\Sigma_{d_1, d_2}$ in order to create a prescribed double curve; the other 
(Algorithm \texttt{UsePinchPoints}), instead, forces the differential of the 
projection to be rank-deficient at the preimages of pinch points. 

Recall that the double curve of~$S$ is irreducible  by assumption. This implies 
that, in presence of pinch points, the curve in~$\Sigma_{d_1, d_2}$ that is 
mapped $2:1$ to the singular curve of~$S$ is also irreducible, because the two 
sheets of the double cover meet precisely at the preimages of the pinch points.
Let $W \subset \p^2$ be the singular image of a good projection $S 
\longrightarrow \p^2$, where $S$ is a good rational ruled surface. Let $W' 
\subset \rho^{-1}(W)$ be the curve that is mapped $2:1$ to the singular 
curve~$Z$ of~$S$. The curve~$W'$ has bidegree 
\[
 \bigl(d_1 + d_2 - 2, (d_1 + d_2 - 2)(d_2 - 1) \bigr) \,.
\]
In fact, by the so-called \emph{double point formula} (see 
\cite[Section~9.3]{Fulton1998} for the general formula, and \cite[Chapter~10, 
Equation~10.52]{Dolgachev2012} for a specialization of the formula in our 
case) its divisor class is given by $(d_1 + d_2 - 2)(H - L)$, where $H$ is the 
class of a hyperplane section of~$\Sigma_{d_1, d_2}$ and $L$ is the class of a 
line; the bidegree of~$W'$ then follows from the fact that $H$ has 
bidegree~$(1, d_2)$ and $L$ has bidegree~$(0,1)$.

For any general point $p \in W'$ there is a unique $q \in W'$ such that 
$\{p,q\}$ is a fiber of the double cover of~$Z$. Note that the 
restriction~$\rho'$ of~$\rho$ to~$W'$ is also a $2:1$ map whose fibers are 
equal to the fibers of $W' \longrightarrow Z$. This allows us to compute~$q$ in 
terms of~$p$ using~$\rho$. We say that $q$ is the \emph{mate} of~$p$.

In order to express the conditions imposed by our algorithm more 
explicitly, we can assume without loss of generality that the 
projection $\p^3 \dashrightarrow \p^2$ is the map forgetting the last 
coordinate. Notice that the map $\Sigma_{d_1, d_2} \longrightarrow \p^3$ 
is defined by four linear forms, three of which we already know from~ the 
map $\rho \colon \Sigma_{d_1, d_2} \longrightarrow \p^2$ obtained via Algorithm 
\texttt{ReconstructRationalScroll}. Let $F_0$, $F_1$, 
and~$F_2$ be the known forms, and let $F$ be the form to be determined. For 
every general point $p \in W'$ and its mate~$q$, we get a linear condition in
the coefficients of~$F$ by requiring that the map $(F_0:F_1:F_2:F)$ 
collapses~$p$ and~$q$, namely by asking that 
\begin{equation*}
 F_i(q) F(p) - F_i(p)F(q) = 0 \quad \text{for every } i \in \{0,1,2\} \, .
\end{equation*}
The forms~$F_0$, $F_1$, and~$F_2$ also satisfy this linear condition, so we 
look for a solution of the linear system that is linearly independent 
from~$F_0$, $F_1$, and~$F_2$.

We claim that the dimension of the solution space for the linear system above is
not bigger than~$4$. Assume, indirectly, that there is a fifth linearly 
independent form $G$ satisfying the linear equations defined above. Then, the 
image of the map $\Sigma_{d_1, d_2} \longrightarrow \p^4$ defined by~$F_0$, 
$F_1$, $F_2$, $F$, and~$G$ is a nondegenerate rational ruled surface with a 
double curve of the same degree as the double curve of~$S$, namely 
$\frac{1}{2}(d-1)(d-2)$. Moreover, the projection from~$\p^4$ to~$\p^3$ is 
birational once restricted to the two surfaces. This contradicts the following 
lemma.

\begin{lemma}
\label{lemma:degree_increase}
Let $f \colon \Sigma_{d_1, d_2} \longrightarrow X \subset \p^3$ be a projection 
of a rational normal scroll satisfying our assumptions, namely $X$ is a good 
rational ruled surface. Suppose that $f$ factors through a 
projection~$\tilde{f}$ to~$\p^4$, namely that $f = \pi \circ \tilde{f}$ where 
$\tilde{f} \colon \Sigma_{d_1, d_2} \longrightarrow \p^4$ and $\pi \colon \p^3 
\dashrightarrow \p^3$. Then $\tilde{X} := \tilde{f}(\Sigma_{d_1, d_2})$ has at 
most isolated singularities.
\end{lemma}

\begin{proof}
Assume by contradiction that $\tilde{X}$ has a singular curve. Consider 
a general plane section~$C$ of~$X$ and its preimage~$\tilde{C} \subset 
\tilde{X}$. Both $\tilde{C}$ and $C$ have the same geometric genus because 
they are birational. By Bertini's theorem, since $X$ is the projection of a 
smooth projective variety, the 
singularities of~$C$ are exactly at the intersection with the 
singular locus~$Z$ of~$X$, and their number is $\deg(Z)$. Since $\deg(Z) = 
\frac{1}{2} (d-1)(d-2)$ (see \cite[Section~6]{Piene2005}), and we know that $C$ 
is rational, it follows from the geometric genus formula (see \cite[Chapter~IV, 
Exercise~1.8(a)]{Hartshorne1977}) that the delta invariant of each of the 
singularities of~$C$ is~$1$. The sum of delta invariants of 
singularities of the space curve~$\tilde{C}$ must be strictly smaller than the 
sum of the delta invariants of the singularities of~$C$. This follows from 
the geometric genus formula, since the arithmetic genus $\frac{1}{2} 
(d-1)(d-2)$ of~$C$ is bigger than the arithmetic genus of~$\tilde{C}$, which 
cannot be greater than $\frac{1}{4} \, d^2 - d + 1$  (see \cite[Chapter~IV, 
Theorem~6.4 and Figure~18]{Hartshorne1977}). It follows that the singular 
locus~$\tilde{Z} \subset \tilde{X}$ must have strictly lower degree than $Z 
\subset X$, because every intersection of~$\tilde{C}$ and~$\tilde{Z}$ 
determines a singularity of~$\tilde{C}$. This implies the image of~$\tilde{Z}$ 
under the projection to~$\p^3$ is a component of~$Z$, but by assumption $Z$ is 
irreducible. This contradiction concludes the proof.
\end{proof}

From an algorithmic point of view, to obtain the desired projection 
$\Sigma_{d_1, d_2} \longrightarrow \p^3$, we need to find a solution~$F$ of the 
infinitely many linear equations above (one for each pair of mates) that is 
linearly independent from~$F_0$, 
$F_1$, and~$F_2$. We could just collect sufficiently many points on~$W$ and 
solve the linear equations arising from them and their mates. However, finding 
points on~$W$ is not trivial. What we do instead is to compute the mate of a 
point with coordinates in a transcendental field extension of the base field 
that is isomorphic to the function field of~$W$. More concretely, the equations 
\begin{equation*}
 F_i(q) F_j(p) - F_i(p) F_j(q) = 0 \quad \text{for every } i,j \in \{0,1,2\}
\end{equation*}
allow one to write the coordinates $(u,v)$ of the mate~$q$ of a point $p = 
(s,t)$ as rational functions of~$s$ and~$t$. This is a consequence of the fact 
that whenever we have a $2:1$ map $C \longrightarrow D$ between two curves, 
then there exists a birational automorphism of~$C$ swapping the two points in 
any fiber. This leads to a single linear equation for~$F$ with coefficients in 
this function field. Using Gr\"obner bases, we can eliminate from this single 
equation the generators of the function field and obtain an equivalent system of 
linear equations with scalar coefficients.

The discussion so far proves the correctness of Algorithm 
\texttt{CollapseMates}.
\begin{algorithm}
\caption{\texttt{CollapseMates}}
\begin{algorithmic}[1]
  \Require A map $r \colon (\C^{\ast})^2 \longrightarrow \p^2$ as in 
Equation~\eqref{eq:toric}, whose branching locus is~$B$, and the singular 
image~$W$ of a good projection $S \longrightarrow \p^2$ with proper 
silhouette~$B$, where $S$ is a good rational ruled surface.
  \Ensure A parametrization of the surface~$S$.
  \Statex 
  \State {\bfseries Compute} the preimage of the singular image~$W$ under~$r$. 
Let $h$ be the polynomial defining such preimage.
  \State {\bfseries Select} a factor~$H$ of~$h$ of bidegree $\bigl( d_1+d_2-2, 
\, (d_1+d_2-2)(d_2-1) \bigr)$.
  \State {\bfseries Construct} the system of equations for the mate $q = 
(u,v)$ of a point $p = (s,t)$. Let $F_0$, $F_1$ and $F_2$ be the components of 
the map~$r$. The equations for mates are
  \begin{align*}
   F_0(u,v) \, F_2(s,t) - F_2(u,v) \, F_0(s,t) &= 0 \\
   F_1(u,v) \, F_2(s,t) - F_2(u,v) \, F_1(s,t) &= 0 \\
   H(u,v) &= 0
  \end{align*}
  \State {\bfseries Write} $u$ and $v$ as rational functions~$U(s,t)$ 
and~$V(s,t)$ using the previous equations by computing a 
Gr\"obner basis with an elimination term order.
  \State {\bfseries Set up} a system of equations for the coefficients of a 
polynomial~$F_3$ of the form $F_{23}(t) + s \, F_{13}(t)$ with $F_{i3}$ of 
degree~$d_i$ with indeterminate coefficients as follows:
  \[
   F_0\bigl(U(s,t),V(s,t)\bigr) \, F_3(s,t) - F_3\bigl(U(s,t),V(s,t)\bigr) \, 
F_0(s,t) = 0
  \]
  \State {\bfseries Solve} the linear system for the coefficients of~$F_3$.
  \State \Return the parametrization $(F_0: F_1: F_2: F_3)$.
\end{algorithmic}
\end{algorithm}

The second algorithm we propose is based on the observation that the fourth 
unknown polynomial~$F$ satisfies particularly simple equations coming from pinch 
points: the Jacobian of the projection $\Sigma_{d_1, d_2} \longrightarrow S$ is 
rank-deficient at the preimages of pinch points, more precisely the tangent line 
of~$W'$ at those points is collapsed to a point by the map defined by 
$(F_0:F_1:F_2:F)$. 

For this algorithm to work, we need to suppose that the images of pinch points 
under the projection $S \longrightarrow \p^2$ are transversal intersections of 
the proper silhouette and the singular image. This is true for projections from 
general centers (see \cite[Proposition~2.1]{Gallet2018} and the discussion 
before for a more thorough analysis).

The algorithm works as follows: for each intersection~$P$ of the proper 
silhouette and the singular image coming from a pinch point, one considers the 
corresponding point $P' \in \Sigma_{d_1, d_2}$ that is sent to~$P$ by $\rho 
\colon \Sigma_{d_1, d_2} \longrightarrow \p^2$. Notice that $P'$ is unique by 
assumption, since $P$ comes from a pinch point. The point $P'$ lies on the 
curve~$W'$ which is mapped to the singular image by~$\rho$. We compute the 
tangent line of~$W'$ at~$P'$, and we impose that the differential of the map 
$(F_0:F_1:F_2:F)$ sends it to zero. In this way, we obtain linear conditions for 
the coefficients of~$F$. Solving these linear system provides the desired 
parametrization of~$S$. Notice that we do not know \emph{a priori} which 
transversal intersections of the proper silhouette and the singular image come
from pinch points, therefore in principle one may have have to repeat the 
previous procedure for all possible subsets of $4(d-3)$ transversal intersections
of proper silhouette and singular image (recall that we suppose from the beginning 
that the number of pinch points is $4(d-3)$). We hence get the following algorithm.

\begin{algorithm}
\caption{\texttt{UsePinchPoints}}
\begin{algorithmic}[1]
  \Require A map $r \colon (\C^{\ast})^2 \longrightarrow \p^2$ as in 
Equation~\eqref{eq:toric}, whose branching locus is~$B$, the singular image 
$W$ of a good projection $S \longrightarrow \p^2$ where $S$ is a good rational 
ruled surface of degree~$d$, and the images in~$\p^2$ of the pinch points of~$S$.
  \Ensure A parametrization of the surface~$S$.
  \Statex 
  \State {\bfseries Pick} $4(d-3)$ transverse intersections of~$B$ and $W$.
  \State {\bfseries Compute} the preimages in $(\C^{\ast})^2$ of these $4(d-3)$ points. 
  \State {\bfseries Compute} the preimage $\tilde{W}'$ in $(\C^{\ast})^2$ of the 
singular image $W$.
  \State {\bfseries Define} a polynomial $F_3$ of the form $F_{32}(t) + s \, 
F_{31}(t)$ with $F_{3i}$ of degree~$d_i$ with indeterminate coefficients.
  \For{each preimage $P'$ of the images of the pinch points}
  \State {\bfseries Compute} a tangent vector of~$\tilde{W}'$ at $P'$.
  \State {\bfseries Add} linear equations for the coefficients of~$F_3$ 
obtained by imposing that the map $(F_0: F_1: F_2: F_3)$ sends the tangent 
vector to zero.
  \EndFor
  \State {\bfseries Solve} the linear system for $F_3$.
  \State \Return the parametrization $(F_0: F_1: F_2: F_3)$.
\end{algorithmic}
\end{algorithm}

We show that the solution space for the linear equations in Step~$8$ is exactly 
four-dimensional, thus proving the correctness of the algorithm. The dimension 
of this solution space is at least~$4$ because we know by hypothesis that there 
is a good projection $\Sigma_{d_1, d_2} \longrightarrow \p^3$ whose components 
satisfy the linear conditions. Proposition~\ref{proposition:correctness} implies 
that the dimension cannot be bigger than~$4$. 

We propose an alternative proof of this fact, based on the torsal lines on the 
ruled surface, which reveals some of the underlying geometry of these surfaces. 
The proof that the dimension cannot be bigger goes in two steps: first, we show 
that lines in~$S$ passing through critical values\footnote{Recall that 
the \emph{critical values} of a differentiable map are the images of the points 
at which the differential of the map does not have maximal rank.} of the 
projection are torsal, and secondly we prove that a surface in~$\p^4$ with too 
many torsal lines must be degenerate.

Recall from Equation~\eqref{eq:toric} that a projection $\alpha \colon 
\Sigma_{d_1, d_2} \longrightarrow \p^4$ can be encoded in two 
vectors of polynomials $Q_1, Q_2 \in \C[t]^5$ via the map
\[
 \bigl( \C^{\ast} \bigr)^2 \longrightarrow \p^4, \qquad 
 (s,t) \mapsto Q_2(t) + s \, Q_1(t)
\]
in such a way that $Q_1(t)$ and $Q_2(t)$ are linearly independent for 
every~$t$. The map~$\alpha$ is singular at the point $(s,t)$ if and only if 
the matrix
\[
\left(
\begin{array}{c|c|c}
 Q_1(t) & Q_2(t) & 
 \frac{\partial Q_2}{\partial t}(t) + s 
 \frac{\partial Q_1}{\partial t}(t)
\end{array}
\right)
\]
has rank at most~$2$.

\begin{lemma}
\label{lemma:singular_torsal}
 Let $P' \in \Sigma_{d_1, d_2}$ and let $\alpha \colon \Sigma_{d_1, d_2} 
\longrightarrow \p^4$ be a projection entering in a commutative diagram of 
projections
\[
 \xymatrix@C=1.5cm{
  & \p^4 \ar@{-->}[d] \\
  \Sigma_{d_1, d_2} \ar[r]_{\beta} \ar[ur]^{\alpha} & \p^3
 }
\]
where the image of~$\beta$ is a good rational ruled surface.
Suppose that $\alpha$ is singular at~$P'$. Then $\alpha(P')$ lies on a torsal 
line of~$\alpha \bigl(\Sigma_{d_1, d_2}\bigr)$.
\end{lemma}
\begin{proof}
 Recalling notation introduced above and the fact that a line is torsal when the 
tangent planes of the ruled surface at each of its points coincide, we have to 
prove that the matrix
\[
\left(
\begin{array}{c|c|c|c}
 Q_1(t_0) & Q_2(t_0) & 
 \frac{\partial Q_1}{\partial t}(t_0) & 
 \frac{\partial Q_2}{\partial t}(t_0)
\end{array}
\right)
\]
has rank~$3$, where $(s_0,t_0)$ are the coordinates of~$P'$. Since $\alpha$ is 
singular at~$P'$, this matrix cannot have rank~$4$. On the other hand, if this 
matrix had rank $2$, then the line in $\alpha \bigl(\Sigma_{d_1, d_2}\bigr)$ 
passing through $\alpha(P')$ would be singular. However, there are no singular 
lines in $\alpha \bigl(\Sigma_{d_1, d_2}\bigr)$, since otherwise also the image 
of~$\beta$ would have singular lines, and this is not allowed by our general 
assumption. 
\end{proof}

\begin{lemma}
\label{lemma:torsal_planar}
 Let $T$ be a rational ruled surface in~$\p^4$ with at least $2 (\deg(T) - 2)$ 
torsal lines. Suppose that there exists a projection $T \longrightarrow S$ 
where $S \subset \p^3$ is a rational ruled surface with at most ordinary 
singularities. Then $T$ is degenerate, namely it is contained in a hyperplane.
\end{lemma}
\begin{proof}
 The surface~$T$ is the image of a projection $\alpha \colon \Sigma_{d_1, d_2} 
\longrightarrow T$ from a rational normal scroll. Let $Q_1, Q_2 \in \C[t]^5$, 
as above, be the two vectors of polynomials of degree~$d_1$ and~$d_2$, 
respectively, encoding the projection~$\alpha$. Torsal lines in~$T$ 
correspond to values $t_0 \in \C$ such that the matrix 
\[
M := 
\left(
\begin{array}{c|c|c|c}
 Q_1(t) & Q_2(t) & 
 \frac{\partial Q_1}{\partial t}(t) & 
 \frac{\partial Q_2}{\partial t}(t)
\end{array}
\right)
\]
has rank~$3$ at~$t_0$. These values are precisely the common zeros of the 
determinants of the submatrices $M_0, \dotsc, M_4$ obtained by removing a row 
from the previous matrix. The degree of these determinants (as polynomials 
in~$t$) is at most $2 \bigl(\deg(T) - 2\bigr)$. In fact, elementary column 
operations transform the previous matrix into
\[
\left(
\begin{array}{c|c|c|c}
 \frac{\partial Q_1}{\partial t}(t) & 
 \frac{\partial Q_2}{\partial t}(t) &
 d_1 Q_1 - t \,
 \frac{\partial Q_1}{\partial t}(t) &
 d_2 Q_2 - t \,
 \frac{\partial Q_2}{\partial t}(t)
\end{array}
\right)
\]
and $\deg(T) = d_1 + d_2$. Hence all the five determinants $\mcal{M}_i := 
\det(M_i)$ are of the form $\lambda_i \, \mcal{M}$ for $\lambda_i \in \C$ and 
$\mcal{M} \in \C[t]$. The kernel of~$M^t$ contains the element
\[
 (\mcal{M}_0, - \mcal{M}_1, \dotsc, \mcal{M}_4) = 
 \mcal{M} (\lambda_0, -\lambda_1, \dotsc, \lambda_4) \, ,
\]
thus $\lambda_0 Q_{10}(t) - \lambda_1 Q_{11}(t) + \dotsb + \lambda_4 Q_{14}(t) 
= 0$ and similarly for~$Q_2(t)$. Hence all the points of the form $Q_2(t) + s \,
Q_1(t)$ are contained in a hyperplane, namely $T$ is degenerate.
\end{proof}

We can now prove that the solution space for the polynomial~$F$ defining the 
map $\Sigma_{d_1, d_2} \longrightarrow \p^3$ is exactly four-dimensional. In 
fact, if it were bigger, we would get a projection $\alpha \colon \Sigma_{d_1, 
d_2} \longrightarrow \p^4$ which is singular at all points~$P'$ that are 
preimages of the transversal intersections of the proper silhouette and the 
singular image coming from pinch points of~$S$. By 
\cite[Section~$6$]{Piene2005}, the surface~$S$ has $2 \bigl( \deg(S) - 2 
\bigr)$ pinch points. By Lemma~\ref{lemma:singular_torsal}, they determine $2 
\bigl( \deg(S) - 2 \bigr)$ torsal lines in $T := \alpha \bigl(\Sigma_{d_1, 
d_2}\bigr)$. Lemma~\ref{lemma:torsal_planar} shows the contradiction.

The discussion so far proves the correctness of Algorithm 
\texttt{UsePinchPoints}.

We can sum up the findings of this section in the following algorithm, which 
reconstructs, up to projective automorphisms preserving the silhouette, a good 
rational ruled surface~$S$ starting from the singular image and the proper 
silhouette of a good projection $S \longrightarrow \p^2$:

\begin{algorithm}
\caption{\texttt{ReconstructRatRuledSurface}}
\begin{algorithmic}[1]
  \Require The singular image~$W$ and the proper silhouette~$B$ of a good 
projection $S \longrightarrow \p^2$ of a good rational ruled surface~$S$.
A map 
$r \colon (\C^{\ast})^2 \longrightarrow \p^2$ as in 
Equation~\eqref{eq:toric}, whose branching locus is~$B$, the singular image 
$W$ of a good projection $S \longrightarrow \p^2$ where $S$ is a good rational 
ruled surface, and the images in~$\p^2$ of the pinch points of~$S$.
  \Ensure A parametrization of the surface~$S$.
  \Statex 
  \State {\bfseries Apply} Algorithm \texttt{ReconstructRationalScroll} to the 
input $B$ and obtain a map $r \colon (\C^{\ast})^2 \longrightarrow \p^2$ as in 
Equation~\eqref{eq:toric}, whose branching locus is~$B$.
  \State {\bfseries Apply} Algorithm \texttt{CollapseMates} or 
\texttt{UsePinchPoints} to the map~$r$ and the curves~$B$ and~$W$ and obtain a 
parametrization of~$S$, up to projective automorphisms preserving~$B$ and~$W$.
  \State \Return the parametrization of~$S$.
\end{algorithmic}
\end{algorithm}

\section{A faster parametrization for the silhouette}
\label{parametrization}

The bottleneck of Algorithm \texttt{ReconstructRationalScroll} is the 
computation of the parametrization of the 
dual of the silhouette. Our situation is quite special: by assumption the 
silhouette admits only nodes and cusps. General-purpose algorithms for 
parametrizing curves (as, for example, the one implemented in Maple), do not 
have the possibility to take into account this special structure of the curve. 
This is why we implement an \emph{ad hoc} procedure for the parametrization of 
the silhouette that uses the fact that we only have nodes and cusps. Although 
the methods used are all known, we believe it could be beneficial for the reader 
to have an overview of this algorithm.

We use the well-known technique of \emph{adjoints} to compute the 
parametrization, see \cite[Section~4.7]{Sendra2008}. Given a planar curve~$C$ 
of degree~$d$ with only nodes and cusps, the linear system of adjoints is given 
by those homogeneous forms of degree~$d-2$ that pass through the singularities 
of~$C$. In order to get the adjoint forms, we have to take the homogeneous 
component of degree~$d-2$ of the radical of the Jacobian ideal of~$C$. One way 
to obtain this radical ideal is the following: consider the discriminant of the 
curve~$C$ along a random projection; this is a bivariate homogeneous 
polynomial whose factors~$H_{\mathrm{nodes}}$ of order~$2$ correspond to nodes 
of~$C$ and whose factors~$H_{\mathrm{cusps}}$ of order~$3$ correspond to cusps 
of~$C$. If we add the form~$H_{\mathrm{nodes}} \cdot H_{\mathrm{cusps}}$ to the 
Jacobian ideal of~$C$, then we get its radical. The image of~$C$ under the 
rational map induced by the linear system of adjoints is a rational normal 
curve~$R_{d-2}$ in~$\p^{d-2}$ of degree~$d-2$. Suppose now that a smooth point 
$P \in C$ is known. Then we get a point $P_{d-2}$ in~$R_{d-2}$, and 
the projection from $P_{d-2}$ maps $R_{d-2}$ to a rational normal 
curve~$R_{d-3}$ in~$\p^{d-3}$ of degree~$d-3$. Since the projection is a map 
between smooth curves, it can be extended also to~$P_{d-2}$, which gets mapped 
to the image of the tangent line~$T_{P_{d-2}} R_{d-2}$ under the projection. In 
this way we obtain a point~$P_{d-3} \in R_{d-3}$, so we can repeat the 
procedure until we land on~$\p^1$:
\[
 \xymatrix@C=1.5cm{
 R_{d-2} \subset \p^{d-2} \ar[r]^-{\pi_{P_{d-2}}} & 
 R_{d-2} \subset \p^{d-3} \ar[r]^-{\pi_{P_{d-3}}} &
 \ldots \ar[r] & \p^1 \\
 C \ar[u] \ar[rrru]_{\varphi}
 }
\]
In this way, we get a map $\varphi \colon C \longrightarrow \p^1$, whose 
inverse is the desired parametrization. Notice that the map~$\varphi$ can be 
computed by selecting those adjoint forms that vanish with multiplicity~$d-3$ 
at~$P$.

This discussion leads to the following algorithm:

\begin{algorithm}
\caption{\texttt{ParametrizeSilhouette}}
\begin{algorithmic}[1]
  \Require A rational curve $C \subset \p^2$ with only nodes and cusps and a 
smooth point $P \in C$.
  \Ensure A parametrization $\psi \colon \p^1 \longrightarrow C$ of~$C$.
  \Statex 
  \State {\bfseries Compute} the radical~$J$ of the Jacobian ideal of~$C$: for 
example, consider the discriminant~$H$ of a general projection of~$C$ on a 
line, factor~$H$ and add to the ideal of derivatives of~$C$ the factors of~$H$ 
of order~$2$ and~$3$.
  \State {\bfseries Let} $L$ be the saturation of the ideal generated by the 
equation of~$C$ and by the $(d-3)^{\text{rd}}$ power of the ideal of the 
point~$P$.
  \State {\bfseries Let} $K := J \cap L$.
  \State {\bfseries Compute} a basis~$\mcal{B}$ of the homogeneous component 
of degree $d-2$ of~$K$.
  \State {\bfseries Compute} the inverse~$\psi$ of the map $C \longrightarrow 
\p^1$ induced by~$\mcal{B}$.
  \State \Return $\psi$.
\end{algorithmic}
\end{algorithm}

We implemented the algorithms in Maple and tested it on a computer with an
Intel I7-5600 processor (1400 MHz). We report the timings in 
Table~\ref{table:timings}.

\begin{table}[ht]
 \caption{The table shows the degree~$d$ of the surface~$S$, the 
degree of the proper silhouette~$B$, its number of nodes~$n$ and of 
cusps~$c$ in the ruled case, and the degree of the nodal curve~$N$, of the 
cuspidal curve~$C$, and the number~$i$ of inflection lines in the tangent 
developable case, and the computing time in CPU seconds. Notice that the 
general algorithm developed in~\cite{Gallet2018} takes $4$s and $130$s
in the cases of ruled surfaces of degree $4$ and $5$, respectively.} 
\begin{tabular}{cccccccccccc}
 \toprule
$d$ & $B$ & $n$ & $c$ & $N$ & $C$ & $i$ & time & type \\
\midrule
$4$ & - & - & - & $6$ & $4$ & $6$ & $2$s & developable &  \\
$5$ & - & - & - & $16$ & $5$ & $9$ & $28$s & developable & \\
$6$ & - & - & - & $30$ & $6$ & $12$ & $145$s & developable &  \\
$4$ & $6$ & $4$ & $6$ & - & - & - & $<1$s & ruled & \\
$5$ & $8$ & $12$ & $9$ & - & - & - & $90$s & ruled & \\
\bottomrule
\end{tabular}
 \label{table:timings}
\end{table}

\end{document}